\documentclass[traditabstract,epsf]{aa}
\usepackage{txfonts}
\usepackage{epsfig}

\newcommand{\teff}{\mbox{$T_{\rm eff}$}}
\newcommand{\logg}{\mbox{$\log g$ }}

\newcommand{\kms}{\mbox{km\,s$^{-1}$ }}
\newcommand{\ms}{\mbox{m\,s$^{-1}$ }}

\begin{document}
\title{WASP-103\,b: a new planet at the edge of tidal disruption\thanks{The photometric and radial velocity time-series used in this work are only available in electronic form at the CDS via anonymous ftp to  cdsarc.u-strasbg.fr (130.79.128.5) or via http://cdsweb.u-strasbg.fr/cgi-bin/qcat?J/A+A/}}

\author{ 
M.~Gillon\inst{1}, D.~R.~Anderson\inst{2}, A.~Collier-Cameron \inst{3},  L.~Delrez\inst{1}, C.~Hellier\inst{2}, E.~Jehin\inst{1}, M.~Lendl\inst{4}, P.~F.~L.~Maxted\inst{2}, F.~Pepe\inst{4}, D.~Pollacco\inst{5}, D.~Queloz\inst{6, 4}, D.~S\'egransan\inst{4}, A.~M.~S.~Smith\inst{7,2}, B.~Smalley\inst{2}, J.~Southworth\inst{2}, A.~H.~M.~J.~Triaud\inst{8,4}, S.~Udry\inst{4},  V.~Van Grootel\inst{1}, R.~G. West\inst{5}}

\offprints{michael.gillon@ulg.ac.be}
\institute{
    $^1$ Institut d'Astrophysique et de G\'eophysique, Universit\'e de Li\`ege, All\'ee du 6 ao\^ut 17, Sart Tilman, Li\`ege 1, Belgium \\
    $^2$ Astrophysics Group, Keele University, Staffordshire, ST5 5BG, UK\\
    $^3$ SUPA, School of Physics and Astronomy, University of St. Andrews, North Haugh, Fife, KY16 9SS, UK\\
    $^4$ Observatoire de Gen\`eve, Universit\'e de Gen\`eve, 51 Chemin des Maillettes, 1290 Sauverny, Switzerland\\  
    $^5$ Department of Physics, University of Warwick, Coventry CV4 7AL, UK\\
    $^6$ Cavendish Laboratory, Department of Physics, University of Cambridge, JJ Thomson Avenue, Cambridge, CB3 0HE, UK\\
    $^7$ N.~Copernicus Astronomical Centre, Polish Academy of Sciences, Bartycka 18, 00-716 Warsaw, Poland\\  
    $^8$ Department of Physics and Kavli Institute for Astrophysics \&\ Space Research, Massachusetts Institute of Technology, Cambridge, MA 02139, USA\\ }   

\date{Received date / accepted date}
\authorrunning{M. Gillon et al.}
\titlerunning{WASP-103\,b: a new planet at the edge of tidal disruption}

 \abstract
  {We report the discovery of  WASP-103\,b, a new ultra-short-period
  planet ($P=22.2$ hr) transiting a  12.1 $V$-magnitude F8-type main-sequence star 
   ($1.22 \pm 0.04$ $M_\odot$, $1.44_{-0.03}^{+0.05}$ $R_\odot$, \teff   = 6110 $\pm$ 160 K). WASP-103\,b
  is  significantly more massive ($1.49 \pm 0.09$ $M_{Jup}$) and larger ($1.53_{-0.07}^{+0.05}$
   $R_{Jup}$) than Jupiter. Its large size and extreme irradiation ($\sim9\times10^9$ erg\,s$^{-1}$\,cm$^{-2}$)
make it an exquisite target for a thorough atmospheric characterization with existing facilities. Furthermore, its
orbital distance is less than 20\% larger than its Roche radius, meaning that it might be significantly 
distorted by tides and might experience mass loss through Roche-lobe overflow. 
It thus represents a new key object for understanding the last stage of the tidal evolution 
of hot Jupiters.
}
   
   \maketitle

\section{Introduction}
Ever since the historical detection of the transits of HD\,209458\,b (Charbonneau et al. 2000,
Henry et al. 2000), the ultra-short-period (a few days, or even less) transiting giant planets  
have played a pivotal role for exoplanetary science. These transiting hot Jupiters 
undergo irradiation orders of magnitude  larger than any solar system planets (Fortney et al. 
2007),  and are also subject  to intense gravitational and magnetic fields  (Correia \& Laskar 
2011, Chang et al. 2010). Their eclipsing configuration provides us with a unique opportunity to study 
  their response to such extreme conditions and  to improve substantially our knowledge 
of planetary structure, chemical composition and physical mechanisms. Notably, it allows us 
to study the last stages of their tidal evolution, provided giant planets close to tidal disruption
are found to be transiting bright stars. 

Our WASP transit survey (Pollacco et al. 2006) actively searches for ultra-short-period giant planets.
 Despite a maximized detection probability (higher transit probability, more frequent 
transits),  planets with a period shorter than 1.2\,d represent only a few percent of the total WASP harvest,
demonstrating their rarity (Hellier et al. 2012). This small subgroup has so far been composed of  WASP-12\,b (Hebb et al. 2009),
WASP-18\,b (Hellier et al. 2009), WASP-19\,b (Hebb et al. 2010), and WASP-43\,b (Hellier et al. 2011a). Among these
four planets, WASP-12\,b and WASP-19\,b are the only two gas giants close to tidal disruption known to transit a bright star.
We report here our discovery of a third planet of this ultra-rare kind, WASP-103\,b,
which orbits a 12.1 $V$-magnitude F-type star at only $\sim$3 stellar radii.

Section~2 presents our discovery and follow-up data. In Sect.~3, we briefly present the spectroscopic determination 
of the stellar properties and the derivation of the system parameters  through a combined analysis of all our data. We discuss our 
discovery in Sect.~4.

\section{Observations}

\subsection{WASP transit detection photometry}
The host star WASP-103 (1SWASPJ163715.59+071100.0 = 2MASS16371556+0711000; $V$=12.1, $K$=10.8) was observed by the southern station of the WASP survey (Hellier et al. 2011b) during the 2010, 2011, and 2012 observing seasons, covering the intervals  2010 May 15 to  Aug 16, 2011 Mar 26 to Aug 20, and 2012 Mar 25 to Jun 28.  The 11565 pipeline-processed photometric measurements were detrended and searched for transits using the methods described by Collier-Cameron et al. (2006). The selection process (Collier-Cameron et al. 2007) elected WASP-103 as a high priority candidate presenting a periodic transit-like signature with a period of 0.926~days.  Fig.~1 (top panel) presents the WASP photometry folded on the best-fit transit ephemeris. 

A search for periodic modulation was  applied to the WASP photometry of 
WASP-103, using the method described in Maxted et al. (2011). No periodic signal was found down to the 3 mmag amplitude.

\begin{figure}
\label{fig:wasp103_1}
\centering                     
\includegraphics[width=9cm]{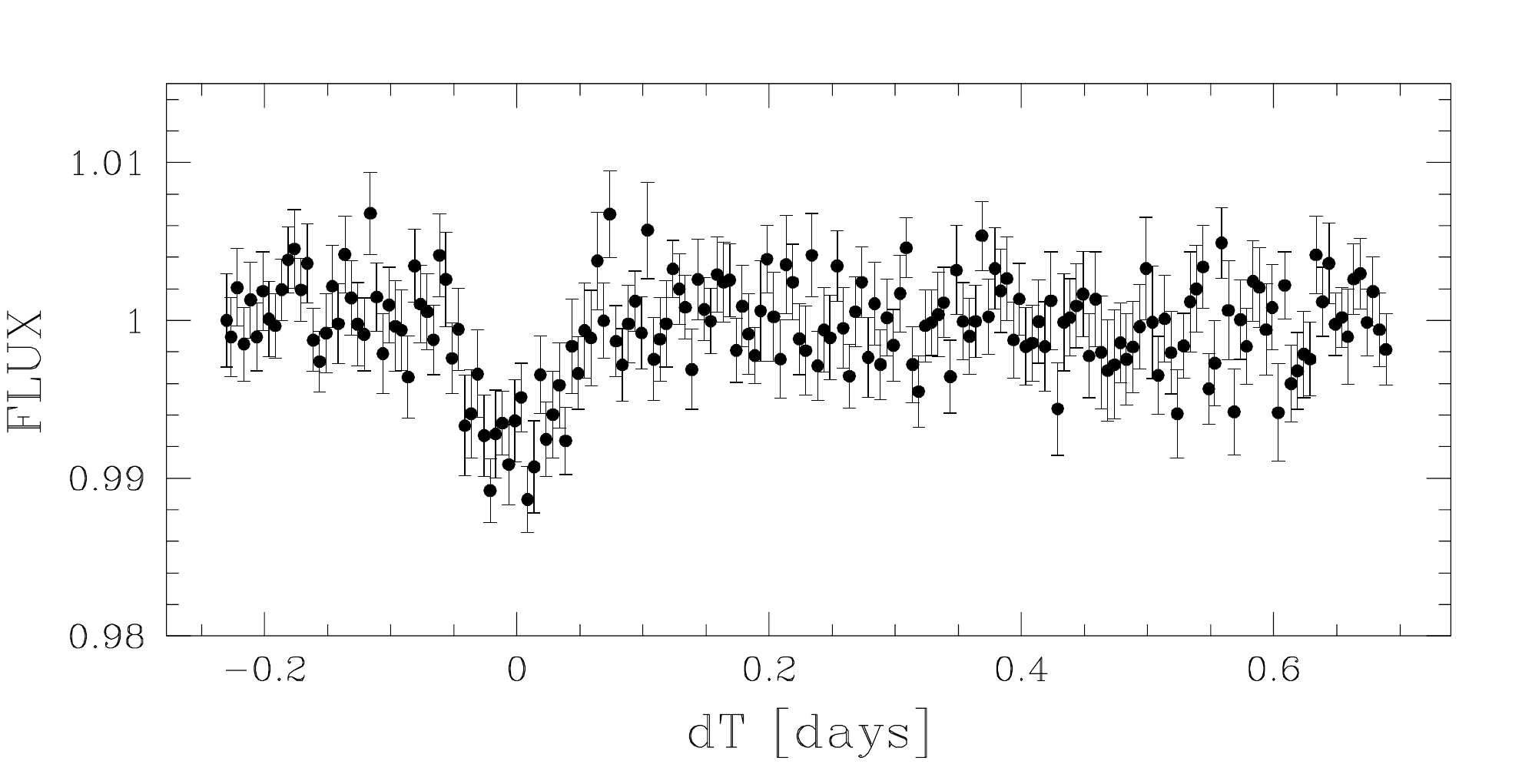}
\caption{WASP photometry for WASP-103
period-folded on the best-fit transit ephemeris from the transit search algorithm presented in Collier-Cameron et al. (2006), 
and binned per 0.005d intervals.}
\end{figure}
 
\subsection{Follow-up photometry}

In 2013, we observed three transits of WASP-103\,b with the robotic 60cm telescope TRAPPIST ({\it TRA}nsiting
{\it P}lanets and {\it P}lanetes{\it I}mals {\it S}mall {\it T}elescope; Jehin et al. 2011) located at the ESO La Silla Observatory 
in the Atacama Desert, Chile.  These transits were observed in a blue-blocking filter\footnote{http://www.astrodon.com/products/filters/exoplanet/} 
that has a transmittance  $>$90\% from 500 nm to beyond 1000 nm. Two transits were also observed  in the Gunn-$r'$ filter with the EulerCam CCD camera at the 1.2-m {\it Euler} Telescope at La Silla Observatory (Lendl et al. 2012). 

For these follow-up eclipse monitorings, the procedures for the observation and data reduction were similar to those described by Gillon et al. (2013, hereafter G13) for WASP-64 and WASP-72, and we refer  to this paper for more details. Table 1 presents a summary of our follow-up photometric time-series obtained for WASP-103. The resulting light curves are shown in Fig. 2.  

\begin{table}
\begin{center}
{\scriptsize
\begin{tabular}{cccccc}
\hline
Night    & Telescope & Filter & $N$         & $T_{exp}$ &  Baseline function  \\ 
              &                    &            &                 &       (s)           &                                  \\ \hline \noalign {\smallskip} 
2013 Jun 15-16 & TRAPPIST &  $BB$ & 802  & 7  &  $p(t^1+f^1)+o$  \\ \noalign {\smallskip} 
2013 Jun 28-29 &{\it Euler} &  Gunn-$r'$ & 103  & 120  &  $p(t^1+f^1+b^2+xy^1)$  \\ \noalign {\smallskip} 
2013 Jul 11-12 &{\it Euler} &  Gunn-$r'$ & 105  & 80  &  $p(t^1+b^1)$  \\ \noalign {\smallskip} 
2013 Jul 23-24 & TRAPPIST &  $BB$ & 941  & 9  &  $p(t^2)+o$   \\ \noalign {\smallskip} 
2013 Aug 4-5 & TRAPPIST &  $BB$ & 935  & 10  &  $p(t^2)+o$   \\ \noalign {\smallskip} 
\hline
\end{tabular}}
\caption{Summary of follow-up transit photometry obtained for WASP-103. $N$=
number of measurements. $T_{exp}$ = exposure time. BB = blue-blocking
filter. The baseline functions are the analytical functions used to model the photometric baseline of each
light curve (see G13).  $o$ denotes an offset fixed at the time of  the meridian flip, and 
$p(\alpha^\beta)$ a $\beta$-order polynomial function of time ($\alpha=t$), full width at half maximum ($\alpha=f$), 
background ($\alpha=b$), or stellar position on the detector ($\alpha=xy$). }
\end{center}
\label{wasp103_trappist}
\end{table}

\begin{figure}
\label{fig:wasp-103_2}
\centering                     
\includegraphics[width=7.5cm]{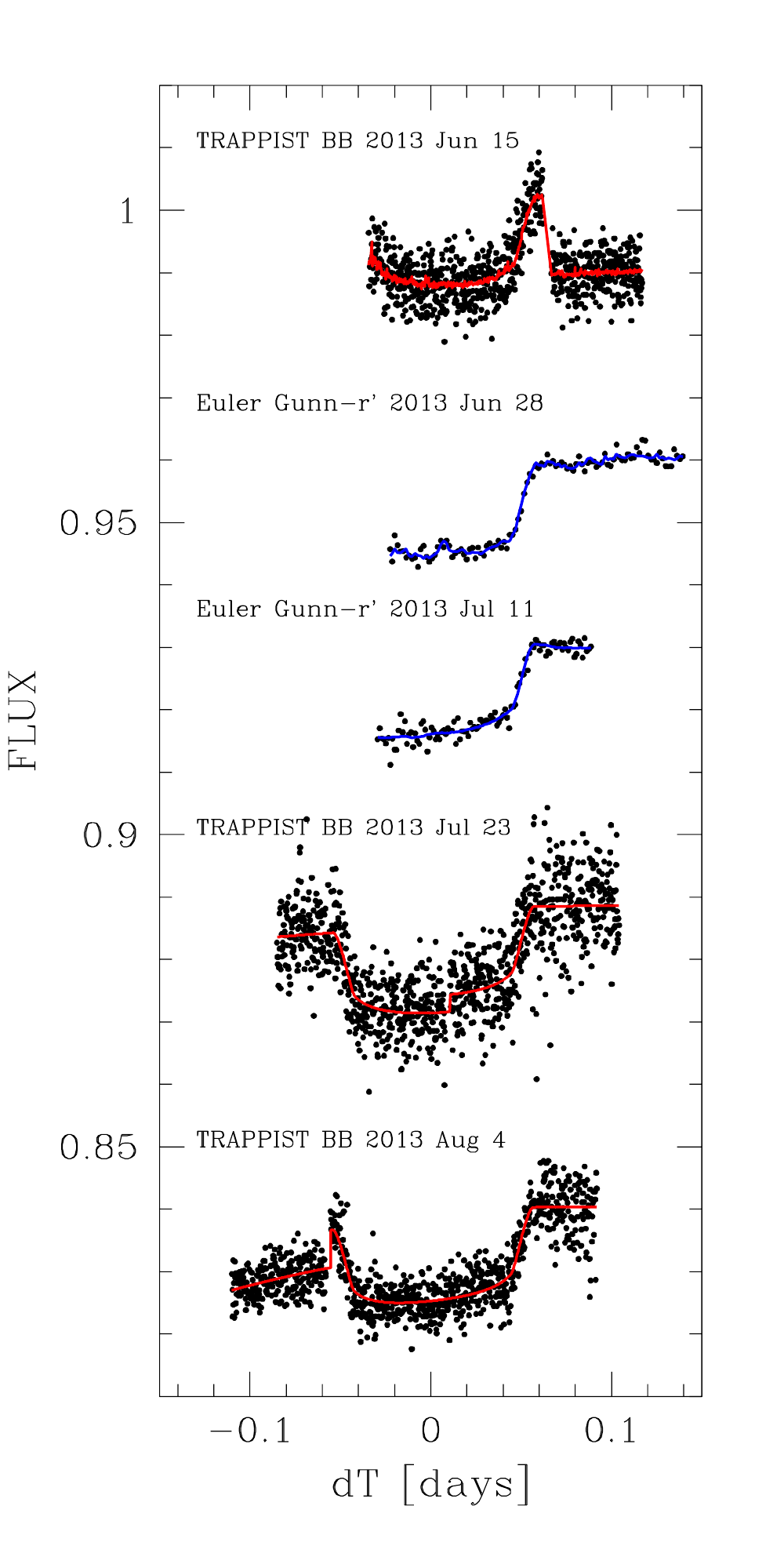}
\includegraphics[width=7.5cm]{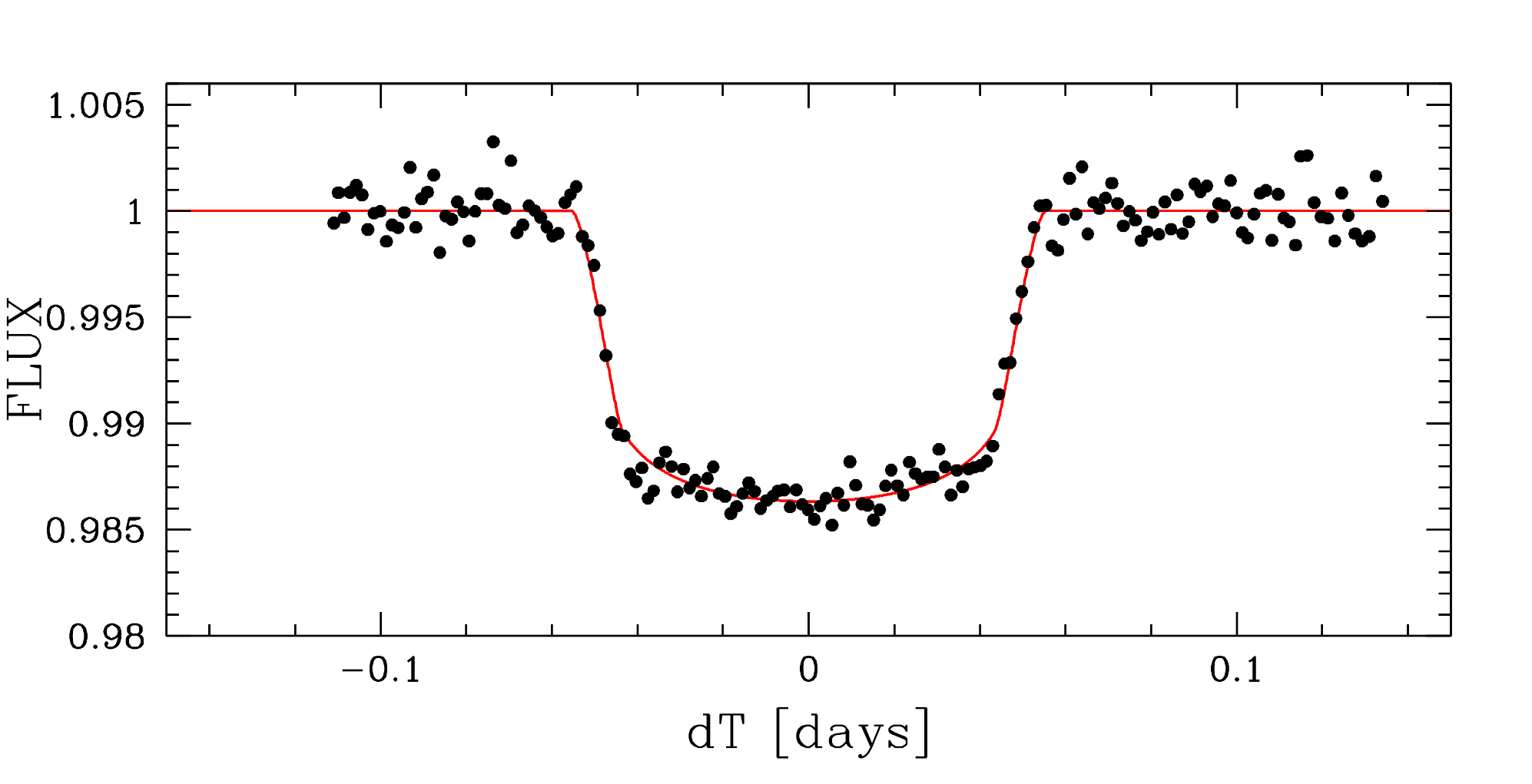}
\caption{$Top$: Follow-up transit photometry for WASP-103\,b.  For 
each light curve, the best-fit transit+baseline model deduced in our global analysis (Sect. 3.2)
 is superimposed. The light curves are shifted along the $y$-axis for clarity.  BB = blue-blocking filter.
The flux offset visible in each BB light curve corresponds to the meridian flip of the TRAPPIST
 German equatorial mount, resulting in a different position of the stars on the chip.
$Bottom$: same follow-up photometry divided by the best-fit baseline functions,  period-folded on the 
best-fit transit ephemeris, and binned in 2 min intervals. Each point corresponds to the 
mean flux value for the corresponding time interval. The best-fit transit model is overplotted in red. }
\end{figure}

\subsection{Spectroscopy and radial velocities}

We gathered 18 spectroscopic measurements with the {\tt CORALIE} spectrograph mounted on {\it Euler} 
to confirm the planetary nature of the body eclipsing WASP-103 and to measure its mass. These spectra were obtained
in the time interval between 2013 Jun 1 to 2013 Sep 10, each
with an exposure time of 30 minutes. Radial velocities  (RVs)  were computed from the spectra by weighted cross-correlation
 (Baranne et al. 1996) with a numerical G2-spectral template that provides close to optimal precisions for late-F to early-K dwarfs, from our 
 experience. 

The RV time-series shows a sinusoidal signal with a period and phase in excellent agreement with those deduced from the WASP
 transit detection, and with an amplitude consistent with a planetary-mass object orbiting a main-sequence star (Fig.~3). 
To confirm that  the RV signal originates  from a planet-mass object orbiting the star, 
we analyzed the {\tt CORALIE} cross-correlation functions (CCF) using the line-bisector technique described in Queloz et al (2001). 
The bisector spans were revealed to be stable, their standard deviation being close to their average error (90 $vs$ 94 \ms), and no evidence for a correlation between the RVs and the bisector spans was found (Fig.~3, lower panel), 
 the slope deduced from linear regression being $-0.06 \pm 0.10$. This bisector analysis  allows 
us to confidently infer that  the RV signal is indeed originating from WASP-103.

\begin{figure}
\label{fig:rv}
\centering                     
\includegraphics[width=8.cm]{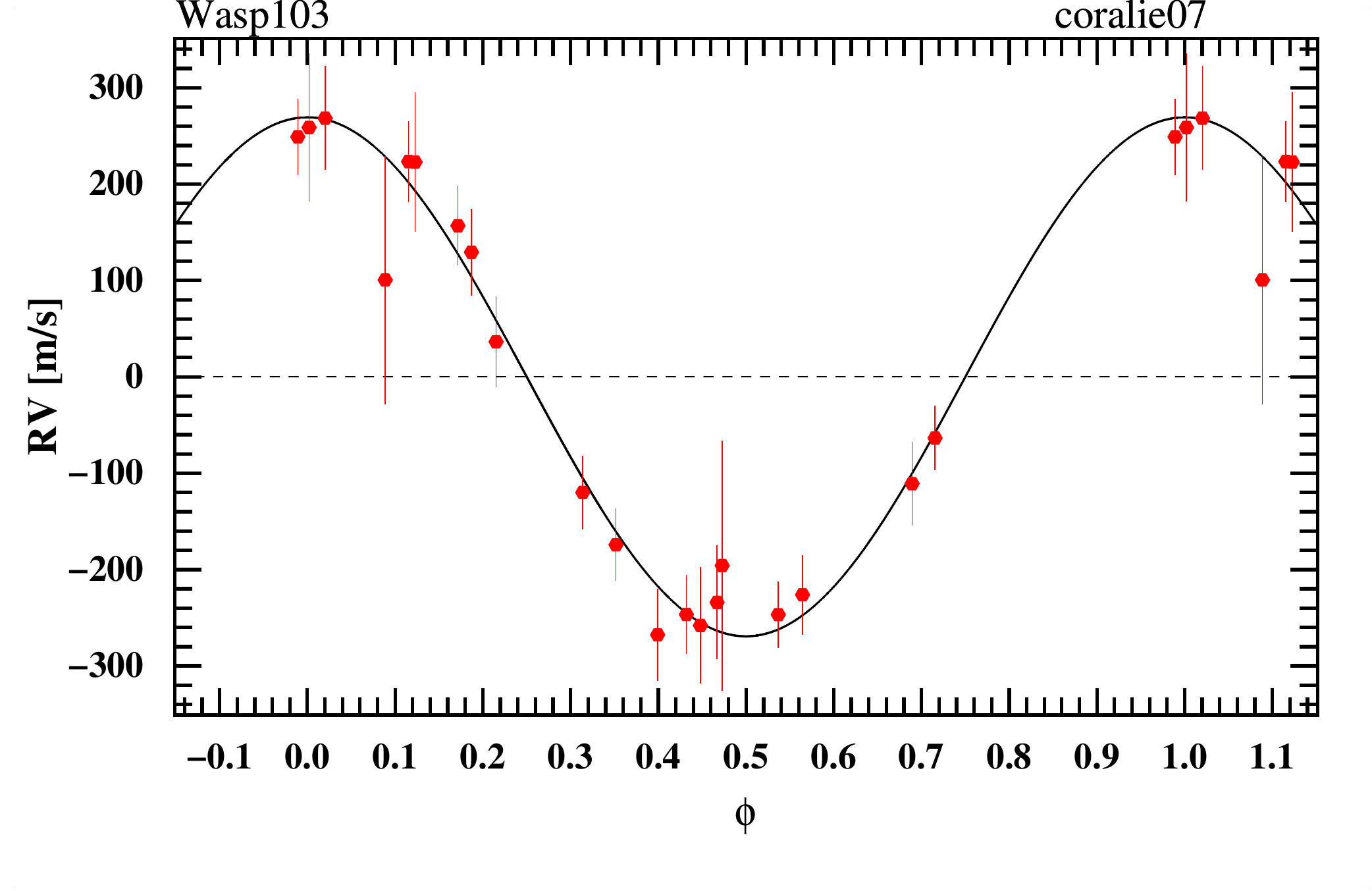}
\includegraphics[width=6.5cm]{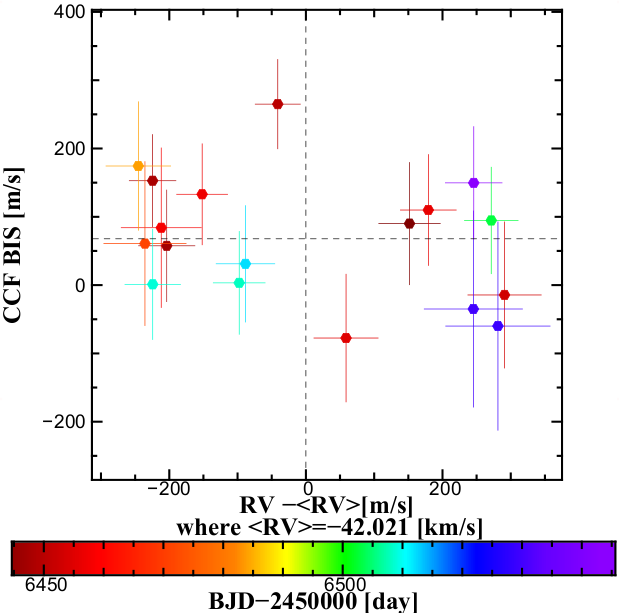}
\caption{{\tt CORALIE} RVs for WASP-103 phase-folded on the best-fit orbital period, and with the best-fit
Keplerian model overplotted. 
$Bottom$: correlation diagram CCF bisector spans $vs$ RV. The colors indicate the measurement timings. 
}
\end{figure}
  
\section{Analysis}

\subsection{Spectroscopic analysis}

The {\tt CORALIE} spectra were co-added to produce a single spectrum with an average signa-to-noise ratio (S/N) of 55. 
A spectral analysis was performed using Fe~{\sc i} and Fe~{\sc ii} equivalent width measurements and 
the methods described by Doyle et al. (2013). The excitation balance of the Fe~{\sc i} lines was used to 
determine the effective temperature (\teff). The surface gravity (\logg) was determined from the ionization 
balance of Fe~{\sc i} and Fe~{\sc ii}. The spectroscopic parameters obtained from the analysis are listed in 
Table 2. The values are quite uncertain because of the low S/N of the co-added spectrum.

\subsection{Global analysis}

We performed a global Bayesian analysis of the follow-up photometry and the
 {\tt CORALIE} RV measurements, probing the posterior probability distribution functions
  of the global model parameters with our adaptive Markov Chain Monte Carlo (MCMC) algorithm 
  (G13 and references therein).
 
 We performed two MCMC analyses, one assuming a
  circular orbit and the other an eccentric orbit. A Bayesian model 
 comparison significantly favored the circular model (Bayes factor $>750$ in its favor), 
 so we selected it as our  fiducial model. The derived values for the system parameters are 
 shown in Table 2.
 
 After  completing  the MCMC analyses,  we checked the values derived for the host star WASP-103 by performing a stellar evolution modeling based on the CLES code (Scuflaire et al. 2008). We derived a stellar mass of 1.18 $\pm$ 0.10 $M_{\odot}$, in excellent agreement with the MCMC result. 
 
\begin{table}
\begin{center}
{\scriptsize
\begin{tabular}{ccc}
\hline \noalign {\smallskip}
Stellar parameters   & Value             & Unit     \\ \noalign {\smallskip}
  \hline \noalign {\smallskip}
RA (J2000)                                                                             &  16 37 15.57             & HH:MM:SS.SS     \\ \noalign {\smallskip} 
DEC (J2000)                                                                          &  +07 11 00.07          & DD:MM:SS.SS     \\ \noalign {\smallskip} 
$V$-magnitude$^b$                                                            & $12.1 \pm 0.01$      &   mag     \\ \noalign {\smallskip} 
$K$-magnitude$^c$                                                           & $10.8 \pm 0.02$      &    mag     \\ \noalign {\smallskip} 
Spectral type                                                                        & F8V                             &               \\ \noalign {\smallskip} 
Effective temperature $T_{eff}$                                        & $6110 \pm 160$        & K            \\ \noalign {\smallskip} 
Metallicity [Fe/H]                                                                  &  $0.06 \pm 0.13$      & dex        \\ \noalign {\smallskip} 
Lithium abundance $\log{A(Li)}$                                     & $2.23 \pm 0.13$        & dex        \\ \noalign {\smallskip} 
Age$^d$                                                                                & 3-5                               & Gyr        \\ \noalign {\smallskip} 
Projected rotation velocity  $V\sin{I_\ast}$                             & $10.6 \pm 0.9$           & \kms      \\ \noalign {\smallskip} 
Microturbulence $V_r$                                                       & $1.1 \pm 0.2$                               & \kms                        \\ \noalign {\smallskip} 
Surface gravity $\log g_*$                                                  & $4.22_{-0.05}^{+0.12}$             & cgs                          \\ \noalign {\smallskip} 
Systemic RV                                                                         &  $-42.001 \pm 0.005$                  & \kms                        \\ \noalign {\smallskip} 
Density $\rho_* $                                                                 & $0.414_{-0.039}^{+0.021}$       & $\rho_\odot $         \\ \noalign {\smallskip} 
Mass $M_\ast $                                                                    & $1.220_{-0.036}^{+0.039}$       &  $M_\odot$             \\ \noalign {\smallskip} 
Radius  $ R_\ast $                                                               & $1.436_{-0.031}^{+0.052}$        & $R_\odot$              \\ \noalign {\smallskip} 
Luminosity $L_\ast$                                                            & $2.59_{-0.32}^{+0.39}$               &  $L_\odot$             \\ \noalign {\smallskip} 
Distance $d$                                                                         & $470 \pm 35$               &  parsec                   \\ \noalign {\smallskip} 
\hline \noalign {\smallskip}
Planet parameters   & Value             & Unit \\ \noalign {\smallskip}
  \hline \noalign {\smallskip}
RV semi-amplitude $K$                                               & $271 \pm 15$                                        & \ms                        \\ \noalign {\smallskip}
Planet-to-star  area ratio  $(R_p/R_\ast)^2 $           & $1.195_{-0.038}^{+0.042}$                & \%                         \\ \noalign {\smallskip}
Planet-to-star radius ratio  $R_p/R_\ast $                & $0.1093_{-0.0017}^{+0.0019}$         &                                \\ \noalign {\smallskip}
Transit impact parameter $b$                                    &  $0.19 \pm 0.13$                                   & $R_*$                    \\ \noalign {\smallskip}
Transit duration  $W$                                                  &   $2.593 \pm 0.024$                              & $hours$                \\ \noalign {\smallskip}
Time of inferior conjunction $T_0$                          &  $2456459.59957 \pm 0.00075$        &  BJD$_{TDB}$     \\ \noalign {\smallskip}
Orbital period  $ P$                                                      &  $0.925542 \pm 0.000019$                &  $d$                         \\ \noalign {\smallskip}
Orbital inclination $i_p$                                              &  $86.3 \pm 2.7$                                     &  $deg$                     \\ \noalign {\smallskip}
Orbital semi-major axis $ a $                                     &  $0.01985 \pm 0.00021$                     &  $AU$                      \\ \noalign {\smallskip}
Orbit scale factor $a / R_\ast$                                    & $2.978_{-0.096}^{+0.050}$                 &                                  \\ \noalign {\smallskip}
Mass  $ M_p$                                                                &  $1.490 \pm 0.088$                               &  $M_{\rm Jup}$     \\ \noalign {\smallskip}
Radius  $ R_p $                                                            &  $1.528_{-0.047}^{+0.073}$                &  $R_{\rm Jup}$      \\ \noalign {\smallskip}
Surface gravity $\log g_p$                                          &  $3.197_{-0.041}^{+0.036}$                &  cgs                          \\ \noalign {\smallskip}
Density  $ \rho_p$                                                        &  $0.550_{-0.070}^{+0.061}$                 & g\,cm$^{-3}$             \\ \noalign {\smallskip}
Equilibrium temperature$^e$ $T_{eq}$                   & $2508_{-70}^{+75}$                              &  K                                 \\ \noalign {\smallskip}
Irradiation                                                                       & $9.1_{-0.8}^{+1.4}$ $10^9$                &  erg\,s$^{-1}$.cm$^{-2}$ \\ \noalign {\smallskip}
Roche limit $a_R$$^f$                                              & $0.01695_{-0.00064}^{+0.00085}$    &  $AU$                        \\ \noalign {\smallskip}
$a/a_R$                                                                         &  $1.162_{-0.052}^{+0.042}$                 &                                    \\ \noalign {\smallskip}
\hline \noalign {\smallskip}
\end{tabular}}
\caption{Value and 1-$\sigma$ errors derived for the WASP-103 star+planet system 
from our analysis. $^a$2MASS. $^b$Egret et al. 1992. $^c$Skrutskie et al 2006. $^d$ from lithium abundance and
stellar evolution modeling. $^e$Assuming a null Bond albedo ($A_B$=0) 
and isotropic reradiation ($f$=1) (Seager et al. 2005). $^f$Using $a_R = 2.46 R_p (M_\ast/M_p)^{1/3}$ (Chandrasekar 1987). 
}
\end{center}
\end{table}

\section{Discussion}

WASP-103\,b is a $\sim$1.5 Jupiter-mass planet orbiting at only $\sim$1.5 stellar diameter of a late F-type star. 
Its size  is significantly larger than predicted by basic models of irradiated planets. For
instance, the models of Fortney et al. (2007) predict a size of $1.25$ $R_{Jup}$  for a core-less $1.5$ $M_{Jup}$ planet
at 0.02 AU from a 1 Gyr solar-twin. The low density of WASP-103\,b is probably related not only to its extreme irradiation close to 
$10^{10}$ erg.s$^{-1}$.cm$^{-2}$, but also to a probably significant level of tidal deformation as its semi-major axis is 
only 15-20\% larger than its Roche radius. These two features make WASP-103\,b join the small subgroup 
of ultra-short-period gas giants suspected to be just at the edge of tidal disruption, which is composed so far of OGLE-TR-56\,b 
(Udalski et al. 2002, Konacki et al. 2003), 
WASP-12\,b,  and WASP-19\,b (Fig. 4). It has been suggested that these extreme planets might experience Roche-lobe overflow, 
mass loss (Erkaev et al. 2007, Li et al. 2010, Lai et al. 2010), and significant tidal distortion (Budaj 2011, Leconte et al. 2011). 
Some signatures of planetary material surrounding WASP-12 have even been detected  by Fossatti et al. (2010, 2013) and  
Haswell et al. (2012). It will thus be extremely interesting to search for signs of mass loss and tidal distortion on WASP-103\,b, 
not only to shed new light on the final stages of the life of hot Jupiters.  We note, however, that the tidal distortion of the planet
probably does not strongly bias the radius measurement presented here. Based on Budaj (2011), we can expect an effective planetary 
radius\footnote{i.e. the radius of the sphere with the same volume as the Roche surface of the planet.} 2-3\% larger than 
measured during the transit, which is significantly smaller than our current error bar ($\sim$5\%). 

The  orbital distribution of hot Jupiters is well explained by assuming an initial formation beyond the ice line followed first by a 
scattering into an eccentric orbit through the influence of a third perturbing body, then by a slow inward migration and orbital 
circularization through tidal dissipation (Ford \& Rasio 2006). In this scenario, most scattered planets
end up on a circularized orbit at 2-4 Roche radii, and from then their orbits evolve only 
through the transfer of angular momentum to the star (e.g. Mastumura et al. 2010); the speed of this final evolution
depends on the mass of the planet and the tidal dissipation efficiency of the star. In this context, the fact that 
OGLE-TR-56\,b, WASP-12\,b, WASP-19\,b and WASP-103\,b have very similar masses (between 1.1 and 1.5 $M_{Jup}$, 
see Fig. 4) may be meaningful,  possibly suggesting a relatively narrow range of tidal dissipation efficiencies for solar-type stars (Chang et al. 2010).

The extreme irradiation of WASP-103\,b, its large size and the brightness of its host star make it a member of the 
select club of transiting planets amenable for a thorough atmospheric characterization with existing ground-based
and space-based facilities (e.g. Seager \& Deming 2010). The comparison of the atmospheric properties of WASP-103\,b 
with those inferred so far for WASP-12\,b (e.g. Sing et al. 2013) and WASP-19\,b (e.g. Anderson et al. 2013) will be particularly useful 
for assessing the common features and peculiarities among the most extreme hot Jupiters. 
 
\begin{figure}
\label{fig:irra_roche}
\centering                     
\includegraphics[width=8cm]{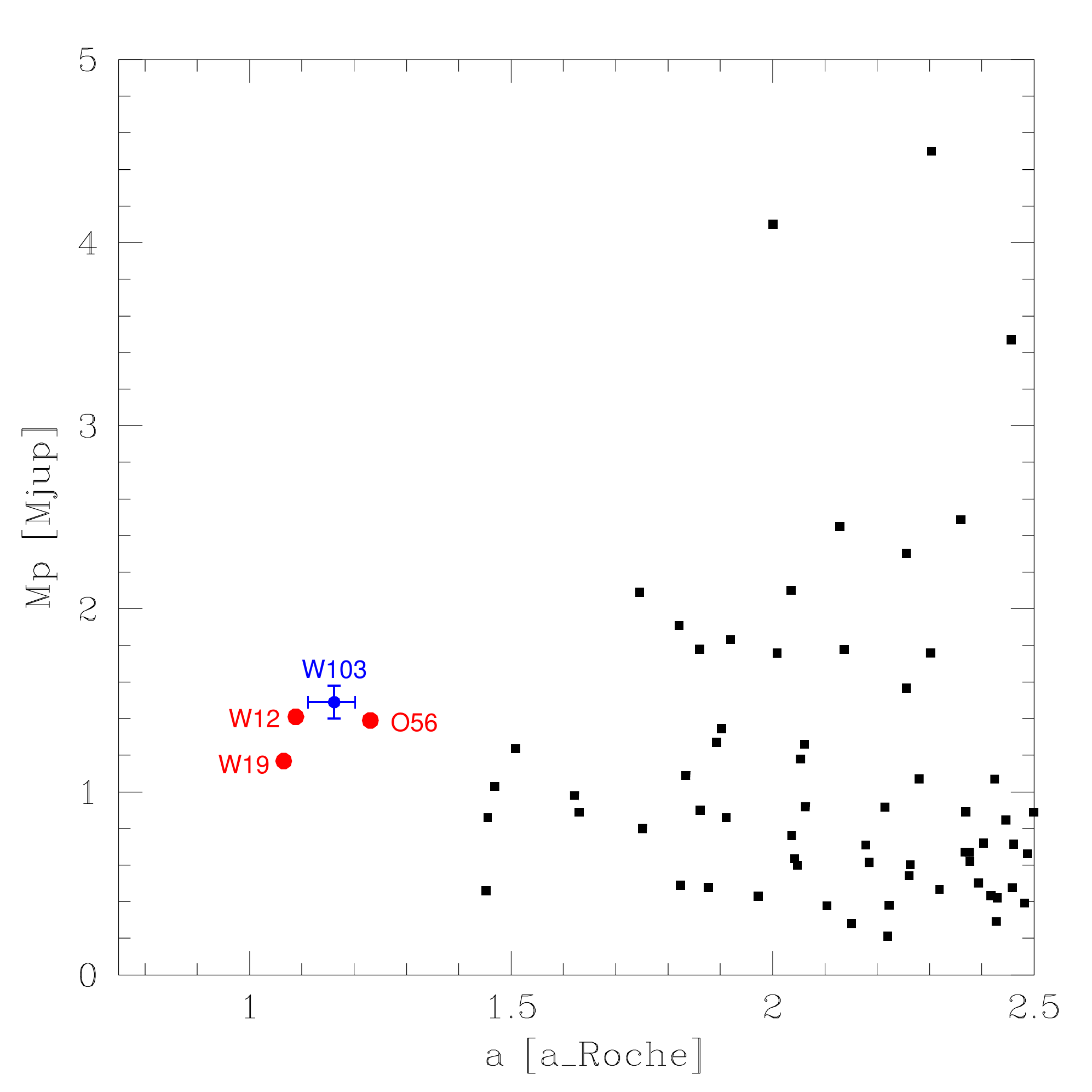}
\caption{Diagram planetary mass $vs$ orbital distance to Roche-limit ratio for known transiting hot Jupiters 
($M>0.2 M_{Jup}$, $P<12$ d; data from NASA Exoplanet Archive). O56=OGLE-TR-56\,b, W12=WASP-12\,b, 
W19=WASP-19\,b, and W103=WASP-103\,b. Here we only show planets with a ratio of the orbital distance to the Roche-limit 
lower than 2.5. }
\end{figure}

\begin{acknowledgements}
WASP-South is hosted by the South African Astronomical Observatory and we are grateful for their ongoing support and assistance. Funding for WASP comes from consortium universities and from UK's Science and Technology Facilities Council. TRAPPIST is a project funded by the Belgian Fund for Scientific Research (Fonds de la Recherche Scientifique, F.R.S-FNRS) under grant FRFC 2.5.594.09.F, with the participation of the Swiss National Science Fundation (SNF).  M. Gillon and E. Jehin are FNRS Research Associates. V. Van Grootel is a FNRS postdoctoral researcher. L. Delrez thanks the Belgian FNRS for funding her PhD thesis. A.~H.~M.~J. Triaud is Swiss National  Science Foundation fellow under grant PBGEP2-145594. This research has made use of the NASA Exoplanet Archive, which is operated by the California Institute of Technology, under contract with the National Aeronautics and Space Administration under the Exoplanet Exploration Program.
\end{acknowledgements}

\end{document}